# An AI-assisted conditioning and geological interpretation workflow for usage in implicit geological modeling

Stefan Carpentier*, Jan Diederik van Wees*, Eva de Boever*, Jan Niederau**, Camille Chapeland*, Suzanne Atkins*, Boris Boullenger*, Jens Wollenweber*

* TNO Netherlands Organisation for Applied Scientific Research,
** Fraunhofer IEG, Fraunhofer Institution for Energy Infrastructures and Geotechnologies, IEG, Aureliusstr. 2, 52062 Aachen, Germany

**Abstract**

Implicit modeling and Relative Geologic Time are geological modeling techniques that enable more efficient, faster, less biased and more reproducible modeling results. For optimal operation, these techniques require many well-constrained input data. In the framework of the Horizon Europe GO-Forward and MOOI WarmingUP GOO projects and to accelerate Implicit modeling, Machine Learning (ML) methods have been tested and implemented in a toolkit for the interpretation of (onshore) seismic data from the shallow to deep range (+- 300 – 3500 m). The goal is to rapidly characterise this depth domain by efficient interpretation of horizons and faults in seismic data. The first step is to improve the signal by applying AI techniques like self-supervised and semi-supervised contrastive learning CNN's for noise reduction and interpolation. Next, horizons and faults are interpreted with minimal use of human-generated training data by using (semi-) self-supervised methods. The resulting developed toolkit supports the application of the implemented algorithms in an efficient workflow. As a first demonstration, the top of the Dutch Maassluis Formation has been interpreted in the Leeuwarden and Waalwijk 3D seismic cubes. Overall, this study demonstrates that AI-assisted interpretation workflows have reached a level of maturity that allows their integration into applied geological modeling and decision-making.

## Introduction

A transition toward implicit geological modeling based on the Relative Geological Time (RGT) paradigm has been pursued within the Geological Survey of the Netherlands (GDN), as an RGT-centric approach has been regarded as a cornerstone for improving subsurface mapping. In implicit modeling, introduced approximately two decades ago by Lajaunie et al. (1997), relative geological time is represented as a three-dimensional scalar potential field that is permitted to be discontinuous at faults and unconformities, such that interpreted horizons are represented as equipotential surfaces. Multiple candidate RGT algorithms, including conventional and AI-based solutions, have been made available and have been shown to connect naturally to complementary model properties (e.g., horizons, faults, and dip fields). Consistency in interpolation and the preservation of full three-dimensional genetic context are imposed by the RGT potential field, and a pathway is provided toward sequence- and chronostratigraphic modeling rather than lithostratigraphic layering. Consequently, the construction of a fully consistent 3D model has been enabled when RGT is constrained with streamlined inputs, including 2D/3D seismic dip information, guiding seismic horizons/units/facies, well tops/markers/formations, and additional geological context.

When it comes to seismic data, the exploration and development of geothermal energy and subsurface High Temperature Storage (ATES) systems face similar challenges:

1. Subsurface data, especially for the layers targeted by ATES and shallow geothermal energy, are often scarce.
2. Many subsurface data have been collected and processed with a focus on hydrocarbon exploration, i.e. deeper layers, or different targets in the subsurface.

These challenges are addressed in the Horizon Europe HE GO Forward project and MOOI project WarmingUP GOO (WarmingUP Geothermal and Storage Scaling), which focus on the derisking, scaling and acceleration of geothermal energy and ATES. In addition to the challenges concerning subsurface data (here with focus on seismic data), geothermal energy and ATES are closely related in the further challenges they face:

3. ATES systems are still in the pilot phase in the Netherlands: scaling up requires gaining experience, developing knowledge, and sharing knowledge.
4. There is more experience with geothermal energy, where better utilization of available data can contribute to increasing the efficiency of individual installations and the sector as a whole.
5. Gaining stronger public support is a point of attention for both geothermal energy and ATES.

To address these challenges, a consistent identification and interpretation of important features, such as horizons and faults in seismic data is of utmost importance. Conventional seismic interpretation and fault detection however use manual picking, tracked picking, and seismic attributes for horizons and faults using routines such as OpendTect auto-tracking and Petrel ant-tracking. This is cumbersome and leads to underutilization of data, as only a limited portion of all data is interpreted/picked in 3D surveys. We propose scaling up the interpretation of seismic data in the Dutch subsurface by interpreting and picking horizons and faults in the full range of data in Dutch 3D surveys instead of just a subset of inlines and crosslines using Machine Learning (ML) algorithms.
Next-generation Machine Learning tools that train and apply neural networks for automated horizon and fault detection have been developed already by market parties. Although generally successful and orders of magnitude faster, there are still many inconsistencies and simplifications in these ML predictions compared to more refined human interpretation/picking. Figure 1 illustrates the general architecture of the AI networks that we employ in our AI-assisted workflow. All of our used networks consist in some way of a Convolutional Neural Network (CNN) and specifically, a U-net which is a customized CNN.

We have developed a ready-to-use AI software tool that uses seismic data and logs as input and generates output on predicted seismic horizons, geological units, and faults. For reducing accuracies and improving customization, we have inspected, conditioned, trained, retrained, and corrected state-of-the-art ML algorithms with the expertise and knowledge of human interpreters and available horizons and faults, such as Dutch Deep/Shallow NLOG datasets (NLOG, 2026). We have cast this into a workflow for simple, user-friendly, low-threshold interaction with the ML tools. The application of the ML tools has been applied on onshore shallow to deep reservoirs/aquifers for ATES and geothermal exploration.
The workflow (Figure 2) starts with AI data conditioning, which performs automatic conditioning of seismic data on large amounts of 2D and 3D seismic data by denoising, interpolation and broadbanding. This is achieved by using Self-supervised learning algorithms ("Noise2Void", Birnie et al., 2021) and a Generative Adversarial Network ("MDA-GAN", Dou et al., 2022, (a)).

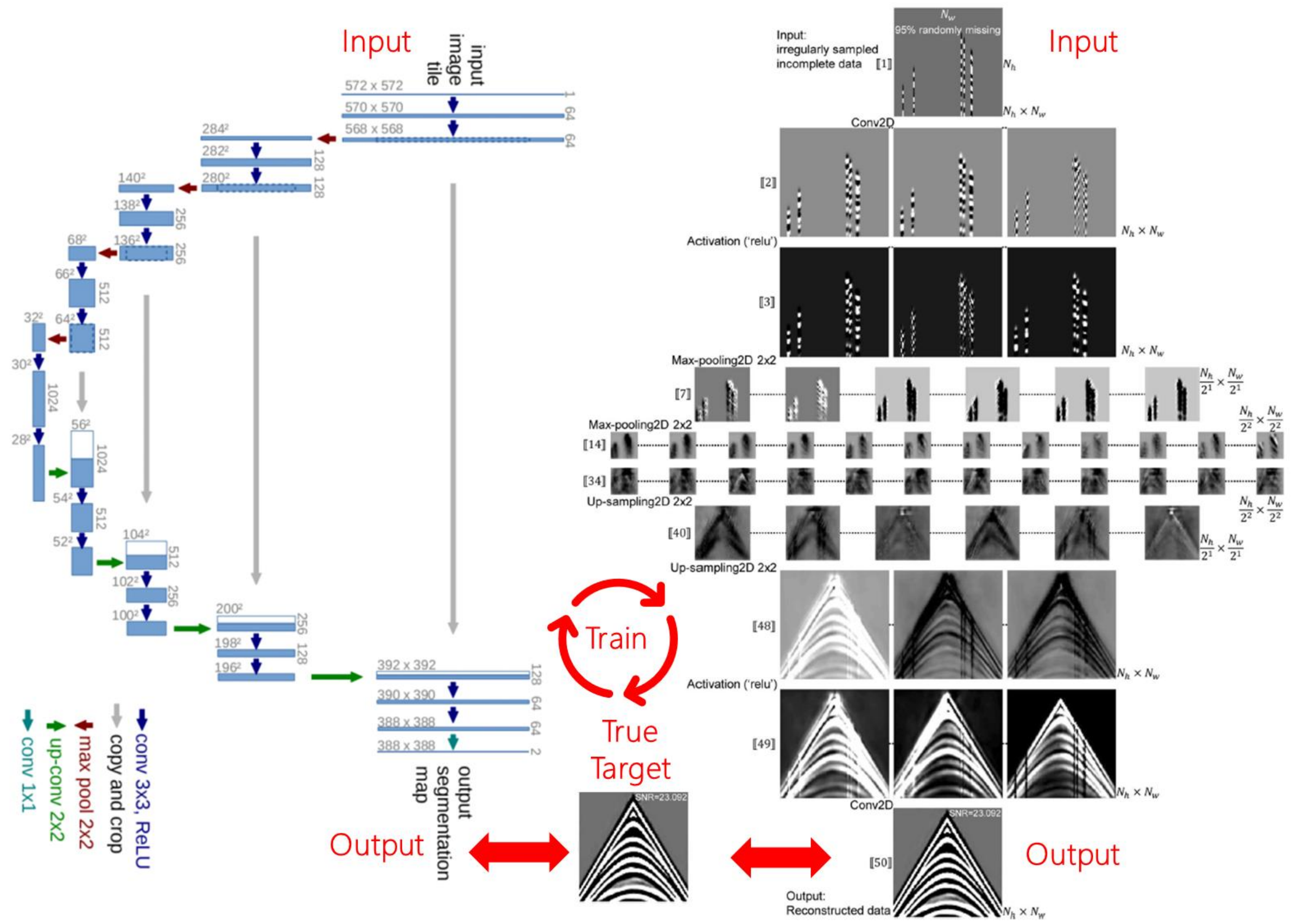

*Fig 1. General architecture of AI neural networks (Ronneberger et al., 2015; Chai et al., 2020). An AI neural network learns the relationship or 'mapping' between the provided Input and the provided True Target. The network is trained by going through iterations where the difference between the Output and True Target is minimized. When the difference becomes minimal, the network is trained and can predict reliable Output based on the provided Input, such as classification of lithology, facies, horizons, faults, etc. The left image shows the schematic architecture of the neural network, and the right image shows the data flow through the neural network. Figure adapted from Carpentier (2025).*

After that, we automatically interpret geological horizons on large amounts of 2D and 3D seismic data. There we employ image segmentation using semi-supervised learning for classifying over- and underburden of given horizons, training on cleaned 3D inlines and crosslines (CONSS, Li et al., 2023). Finally, we automatically interpret faults on large amounts of 2D and 3D seismic data. This is done by image segmentation with semi-supervised learning networks "FaultNet" and "FaultSSL" (Dou et al., 2022 (b), Dou et al., 2023) for fault detection in seismic data, training on Dutch 3D seismic cubes.
Figure 2 illustrates schematically the AI-assisted conditioning and geological interpretation workflow.

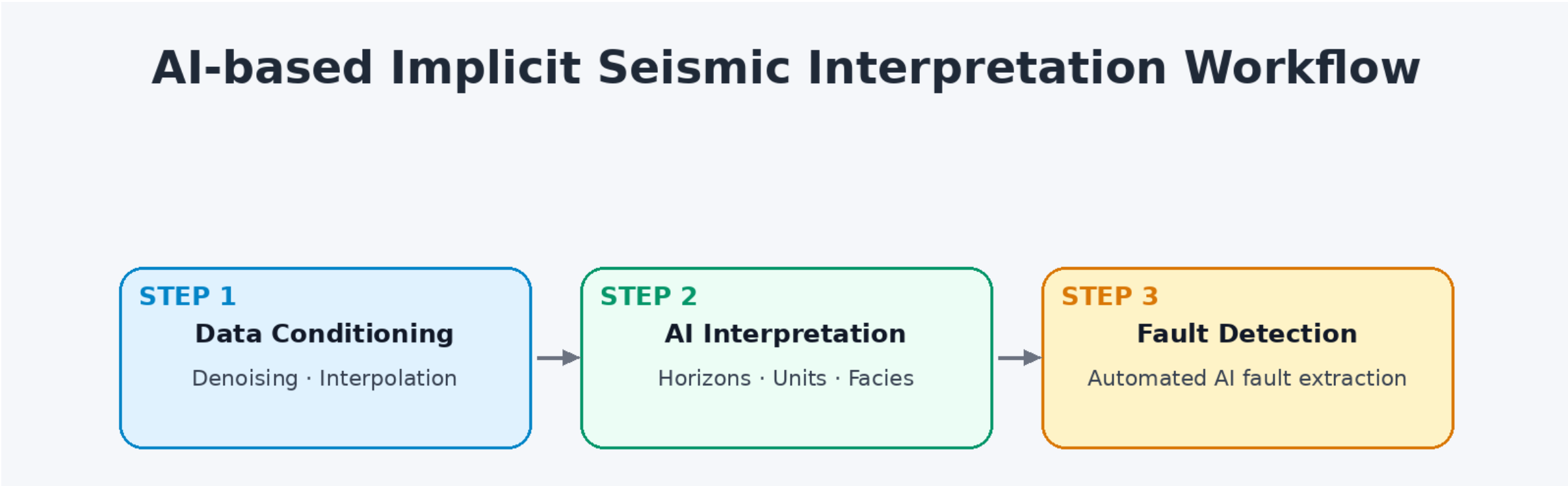

*Fig 2. Schematic overview of the AI-assisted conditioning and geological interpretation workflow. Step 1: Innovative AI based automatized data conditioning (denoising, interpolation) of 2D/3D seismic (dips). Step 2: AI-assisted automatized high resolution interpretation of 2D and 3D seismic, with focus on: horizons, units, facies. Step 3: Apply automatized AI-based seismic fault detection.*

This paper demonstrates the application of the AI-assisted workflow aimed at improving the characterization of the subsurface for ATES and geothermal energy. Specifically, it involves innovative (mostly AI-based) automated workflows for higher resolution interpretations of existing 3D and 2D seismic data, resulting in improved characterization of layers and fault density.

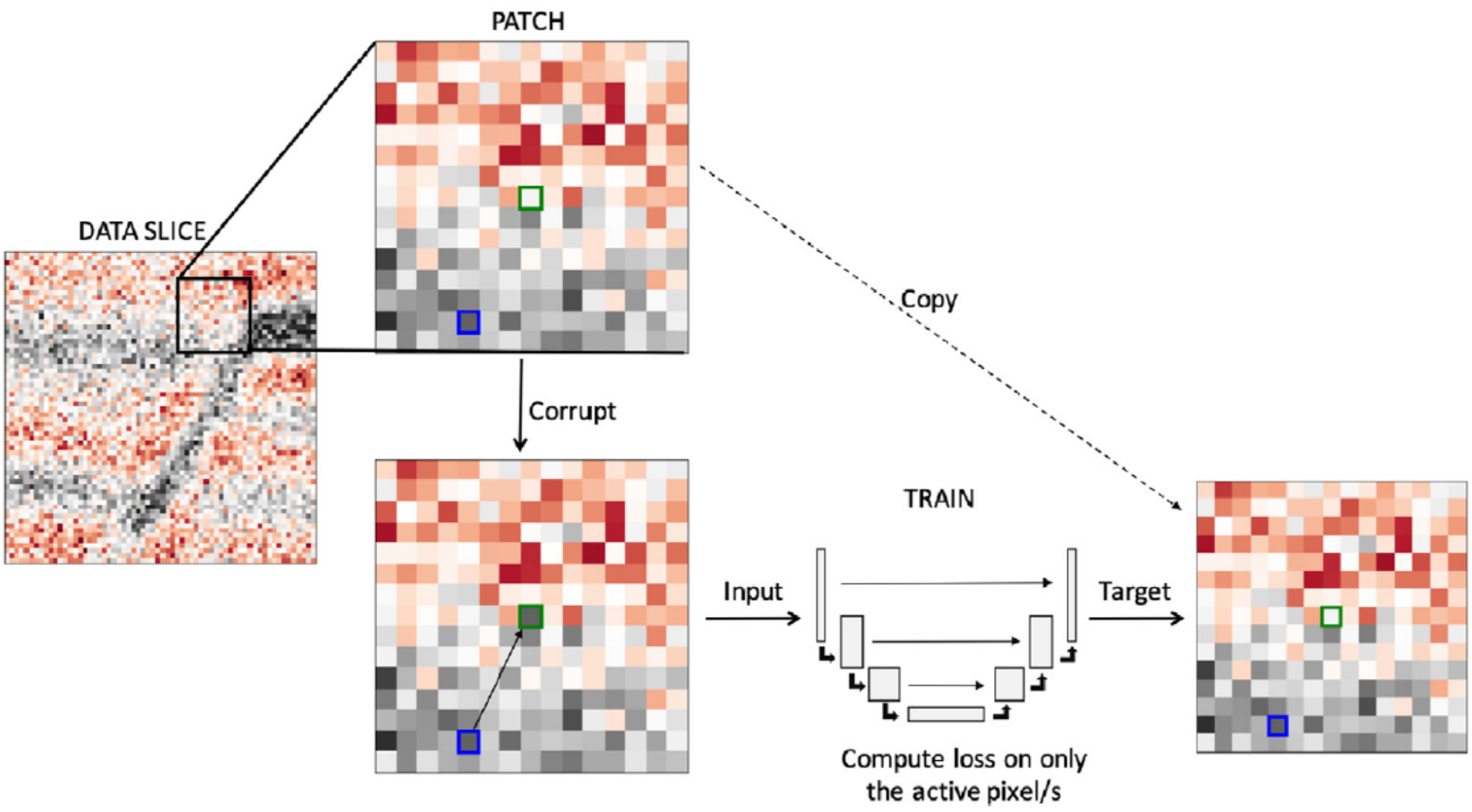

**Fig. 1.** Schematic illustration of the Noise2Void denoising procedure.

*Fig 3. Diagram of AI tool neural network Self-Supervised Denoising "Noise2Void" (Birnie et al., 2021). A Self-Supervised Denoising neural network learns the relationship or 'mapping' between the provided corrupted noisy Patch and the provided original noisy Patch. The network is trained by going through iterations where the network cannot reproduce the unpredictable corrupted pixel in the corrupted noisy Patch but can reproduce the predictable background signal, thereby effectively denoising the input noisy Patch. Figure adapted from Carpentier (2025).*

## Method

2.1 AI seismic data conditioning

A research strategy for AI seismic data conditioning was developed to address the challenge of automatically improving the quality of large volumes of vintage 2D and 3D land seismic data for geothermal exploration. Self-supervised denoising using a blind-spot network (Noise2Void; Birnie et al., 2021) was combined with generative-adversarial interpolation to reconstruct outliers and acquisition gaps (MDA-GAN; Dou et al., 2022a), such that signal continuity was enhanced and random noise was suppressed while reflector character was preserved. In the pilot 3D seismic cube, increased reflector coherence was obtained and more clearly defined Maassluis and Oosterhout reflections were produced, thereby providing an improved basis for subsequent horizon and fault interpretation.

Noise suppression is an essential step in many seismic processing workflows. Part of this noise, especially in land datasets, manifests as random noise. In recent years, neural networks have been successfully used to denoise seismic data in a controlled manner. Controlled AI learning, however, always comes with the often-impossible requirement of having noise-free datasets for training. With AI blind-spot networks, we redefine the denoising task as a self-supervised procedure where the network uses the surrounding seismic noise to estimate the noise-free value of a central sample. Based on the assumption that noise is statistically independent between samples, the network can predict the noise component of the sample due to randomness, while the signal component is accurately predicted due to spatio-temporal coherence. Illustrated on synthetic examples, the blind-spot network proves to be an efficient denoiser of seismic data contaminated by random noise with minimal damage to the signal, providing improvements in both, the image domain and later tasks, such as post-stack inversion. In this study, we use the denoising and generative-adversarial interpolation methods of Birnie et al. (2021), illustrated in Figure 3, and Dou et al. (2022), illustrated in Figure 4.

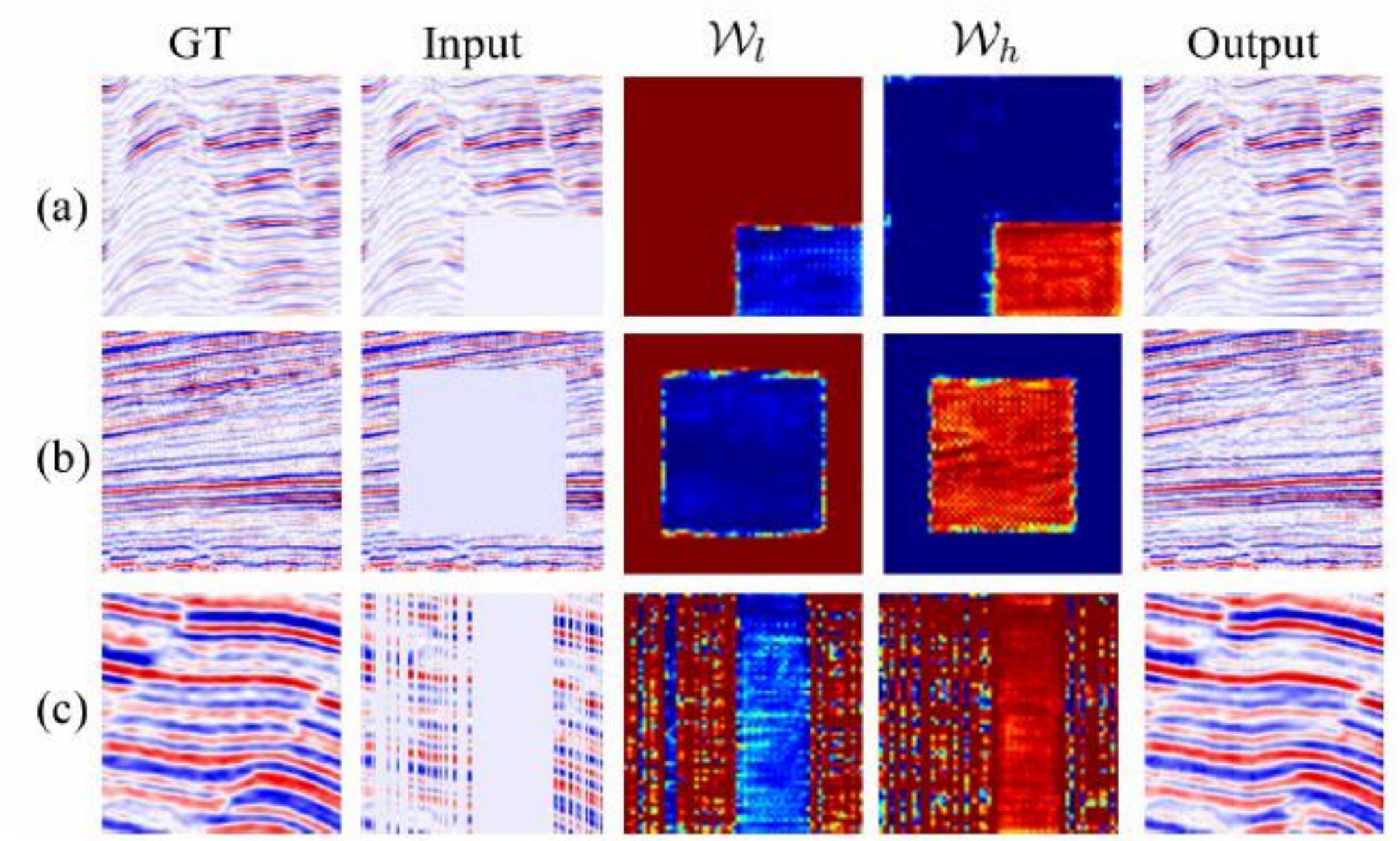

*Fig 4. Concept of AI tool neural network Generative Adversarial Network (GAN) interpolation "MDA-GAN" (Dou et al., 2022, (a)). A conditional Generative Adversarial Network (GAN) learns the relationship or 'mapping' between the provided input seismic data with gaps (Input) and the provided original seismic data without gaps (GT). The network is trained by going through iterations where the network fills the gaps in the seismic data with increasingly realistic 3D complex geological patterns that match the 3D contextual structure of the surrounding undisturbed seismic traces. Figure adapted from Carpentier (2025).*

## 2.2 AI-assisted seismic data interpretation

AI-based horizon interpretation should facilitate the automated interpretation of geological horizons in large volumes of vintage 2D and 3D seismic data for geothermal exploration. Image segmentation based on semi-supervised learning was employed to classify the overburden and underburden of target horizons using CONSS (Li et al., 2023), with training performed on conditioned 3D inlines and crosslines. This is achieved using only 1% of the inlines or crosslines of the cube, taking care to select lines that are representative of the lithology variations in the dataset. In a pilot 3D seismic cube, the Maassluis and Oosterhout horizons were successfully predicted across all inlines and crosslines, thereby demonstrating the feasibility of scalable horizon interpretation with limited reliance on human-labeled data.

The next step is AI interpretation of the cleaned seismic data. Recently, seismic interpretation based on convolutional neural networks (CNN) has made significant progress. However, existing CNN-based AI supervised learning methods require enormous amounts of labeled data. Labeling is labor-intensive and time-consuming, especially for 3D seismic data volumes. To address these issues, Li et al. (2023) propose a semi-supervised method based on contrastive learning, called CONSS, which can efficiently identify geological units with only 1 % of the original human interpretations. Figure 5 shows the diagram of this AI tool. Experimental results in the study by Li et al. (2023) demonstrate that this approach achieves state-of-the-art performance on the Dutch public F3 seismic survey (NLOG, 2026).

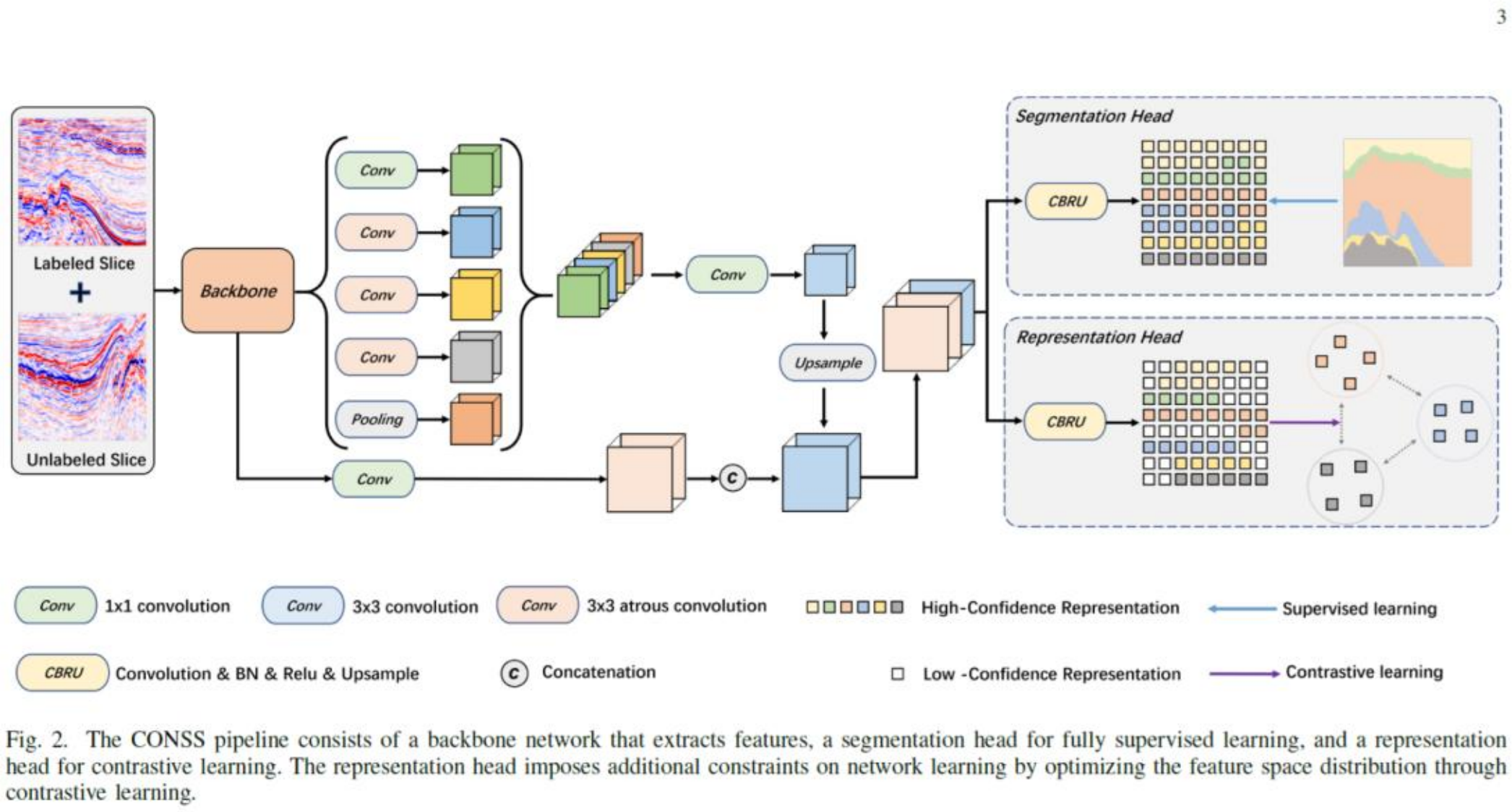

*Fig 5. Diagram of AI tool neural network CONSS (Li et al., 2023). A semi-supervised Contrastive Learning AI neural network learns the relationship or 'mapping' between the provided Input and the provided True Target. The network is first trained to a pretrained model with a lot of general seismic data, to make predictions on new seismic data and retrain the pretrained model to a new model with Contrastive Learning. This way, a general AI neural network is specifically trained on a new dataset with little required training data. Figure adapted from Carpentier (2025).*

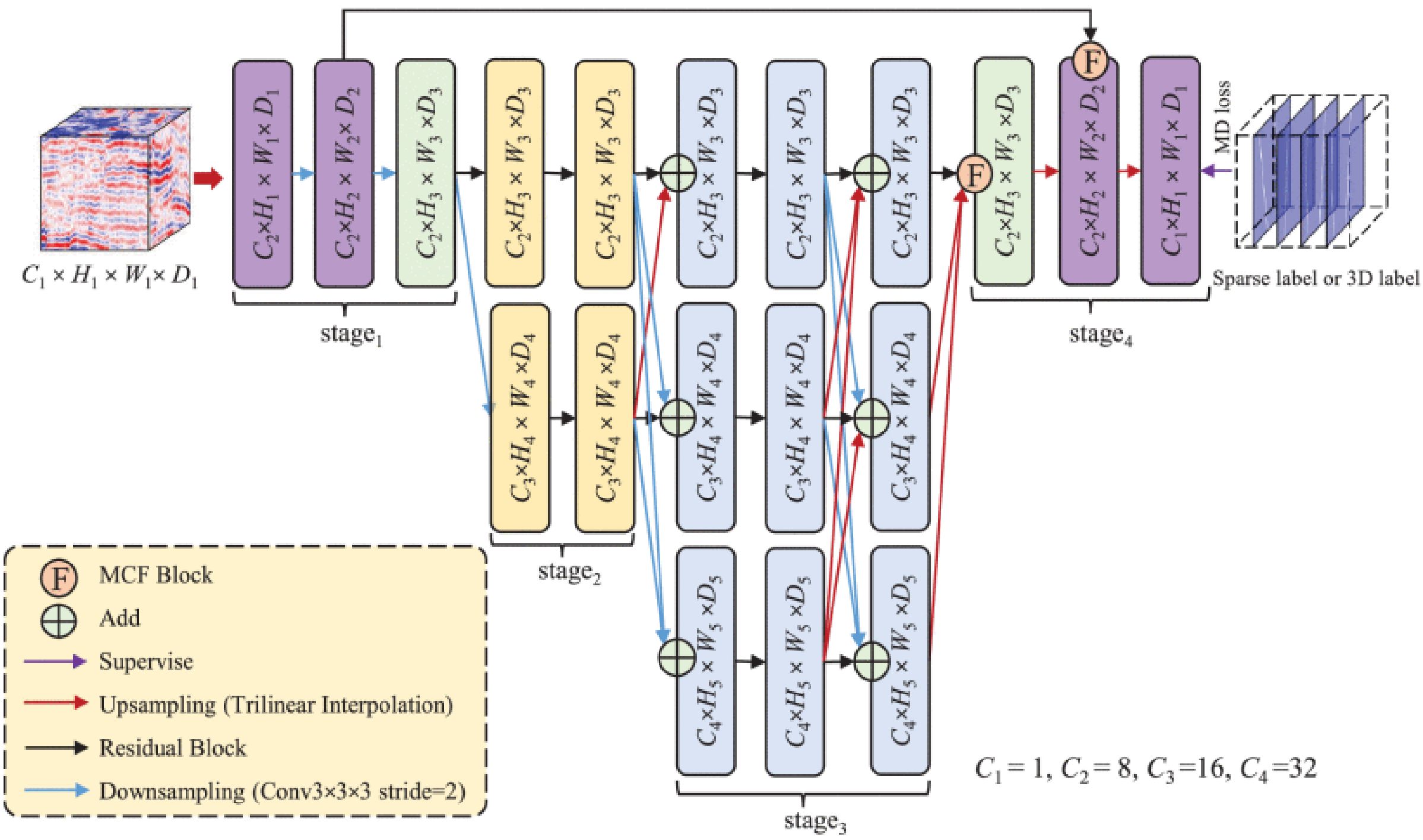

*Fig 6. Diagram of AI tool neural network FaultNet (Dou et al., 2022 (b)). The network is a U-net autoencoder*

*with special weighted labels to properly handle faults parallel to Inlines and Crosslines. Figure adapted from Carpentier (2025).*

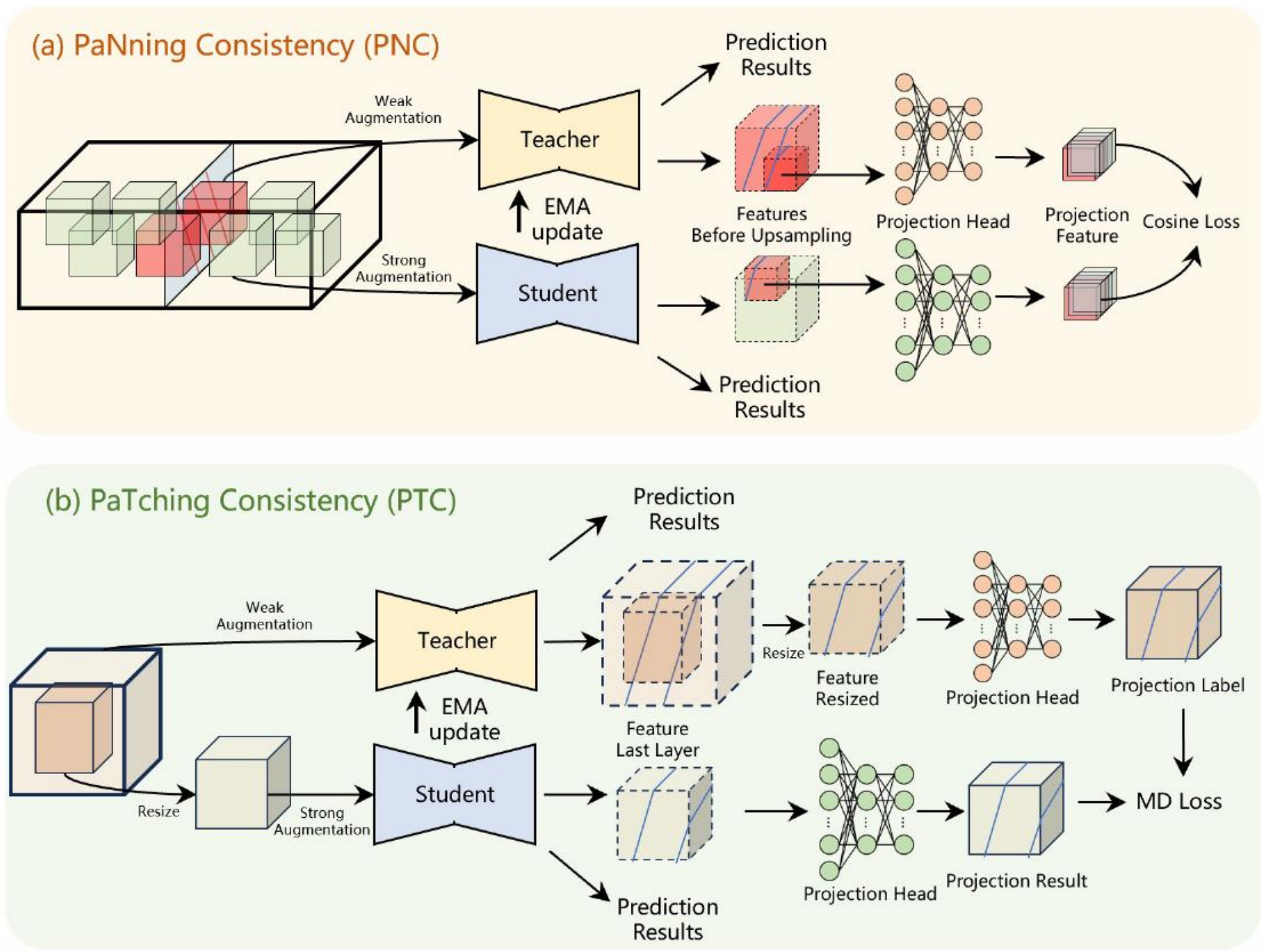

Fig. 1. FaultSSL's PNC and PTC proxy task flowchart.

*Fig 7. Diagram of AI tool neural network FaultSSL (Dou et al., 2023). A semi-supervised AI neural network learns the relationship or 'mapping' between the provided Input and the provided True Target. The network is trained by having a pretrained Teacher model, trained with general seismic data, make predictions on new seismic data and retrain the Teacher model to a Student model with the most reliable predictions of the Teacher model as training data. This way, a general AI neural network is specifically trained on a new dataset with little required training data. Figure adapted from Carpentier (2025).*

## 2.3 AI seismic fault detection

AI-based fault detection is needed to automate the interpretation of faults in large volumes of vintage 2D and 3D seismic data for geothermal exploration. Image segmentation using the semi-supervised learning network FaultSSL (Dou et al., 2022b; Dou et al., 2023) was applied for fault detection, with training performed on the original seismic cube. In a pilot 3D seismic cube, faults were successfully predicted within the Maassluis and Oosterhout formations across all inlines and crosslines, thereby demonstrating the feasibility of scalable fault interpretation with limited labelled data.

Fault detection is a crucial step in energy exploration. Traditional methods rely on seismic attributes and other processing and imaging techniques, which are sensitive to processing parameters and noise. This work focuses on AI methods. The difficulty of obtaining accurate labels has always been a serious problem in data-driven fault detection. The complexity of the 3D distribution of faults makes it almost impossible to manually obtain complete seismic fault labels. 2D labels are relatively easy to obtain, but 2D fault detection has inherent flaws, such as the discontinuity of 2D interpretations in three-dimensional space, the inability to obtain sufficient spatial and contextual information, etc.

To improve the generalization of models trained on limited synthetic datasets to a broader range of field seismic data, Dou et al. (2023) introduce FaultSSL, a semi-supervised AI fault detection method. This method is based on the classic teacher-student structure, where the supervised part uses synthetic data and a set of 2D labels. The unsupervised part relies on two carefully designed proxy tasks, allowing it to incorporate large amounts of unlabelled field seismic data into the training process. Figures 6 and 7 show the architecture of the AI network by Dou et al. (2022 (b)) and Dou et al. (2023).
Custom improvements were implemented by TNO-GDN to enhance the performance of the pre-trained FaultNet and FaultSSL algorithms: 1) The input seismic volume was resampled to 12.5 m × 12.5 m CDP spacing with a 4 ms sample rate to match the training-data sampling, 2) prediction was performed using a default tile size of 256 × 256 × 256 samples. 3) To reduce tile-edge artefacts and improve continuity, overlapping tiles were predicted with blending over a diagonal half-tile size, 4) and post-processing was applied using a time-direction mean filter to suppress purely vertical faults at cube edges, as well as 5) a one-dimensional median filter along the inline and crossline directions. 6) Finally, the fault-probability volume was binarized using a threshold of 0.5, with values < 0.5 set to 0 and values > 0.5 set to 1.

3. Results

Figures 8, 9, and 10 show how the AI self-supervised denoised method of Birnie et al. (2021) successfully suppresses noise without damaging the signal in the example 3D survey "Leeuwarden" (Figures 8 and 9), and how the AI Generative Adversarial Network (GAN) interpolation method of Dou et al. (2022, (a)) improves the continuity of shallow reflectors (Figure 10). The same goes for application on the example 3D survey "Waalwijk" in Figures 11 and 12, where the original data in Figure 11 contains many gaps and high noise levels in the shallow part (0 - 1100 ms) which can be reconstructed from 3D surrounding contextual information by the AI-denoising-interpolation algorithms in Figure 12.

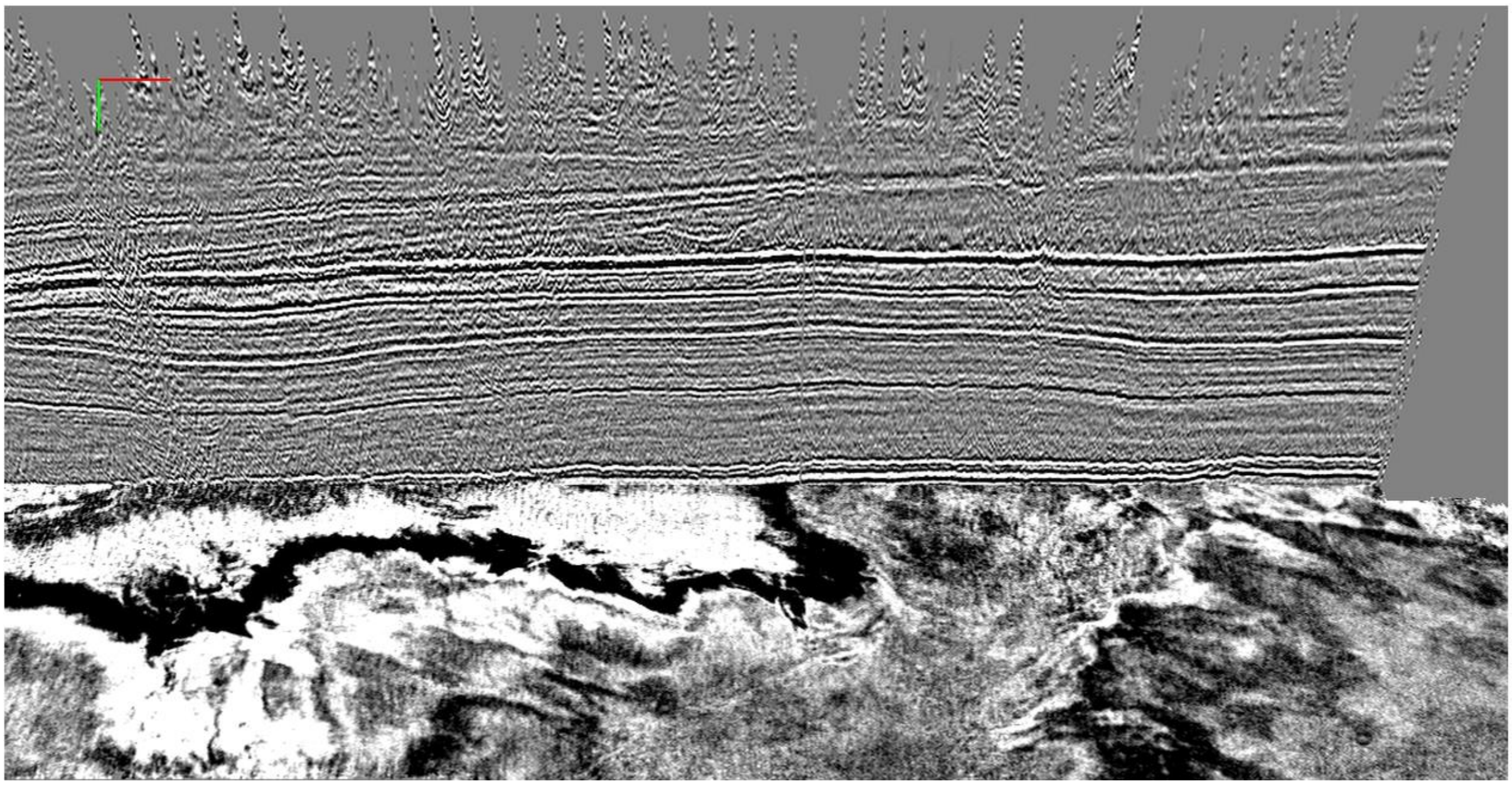

*Fig 8. Dutch onshore 3D seismic cube "Leeuwarden", original northernmost full inline and timeslice 1200 ms.*

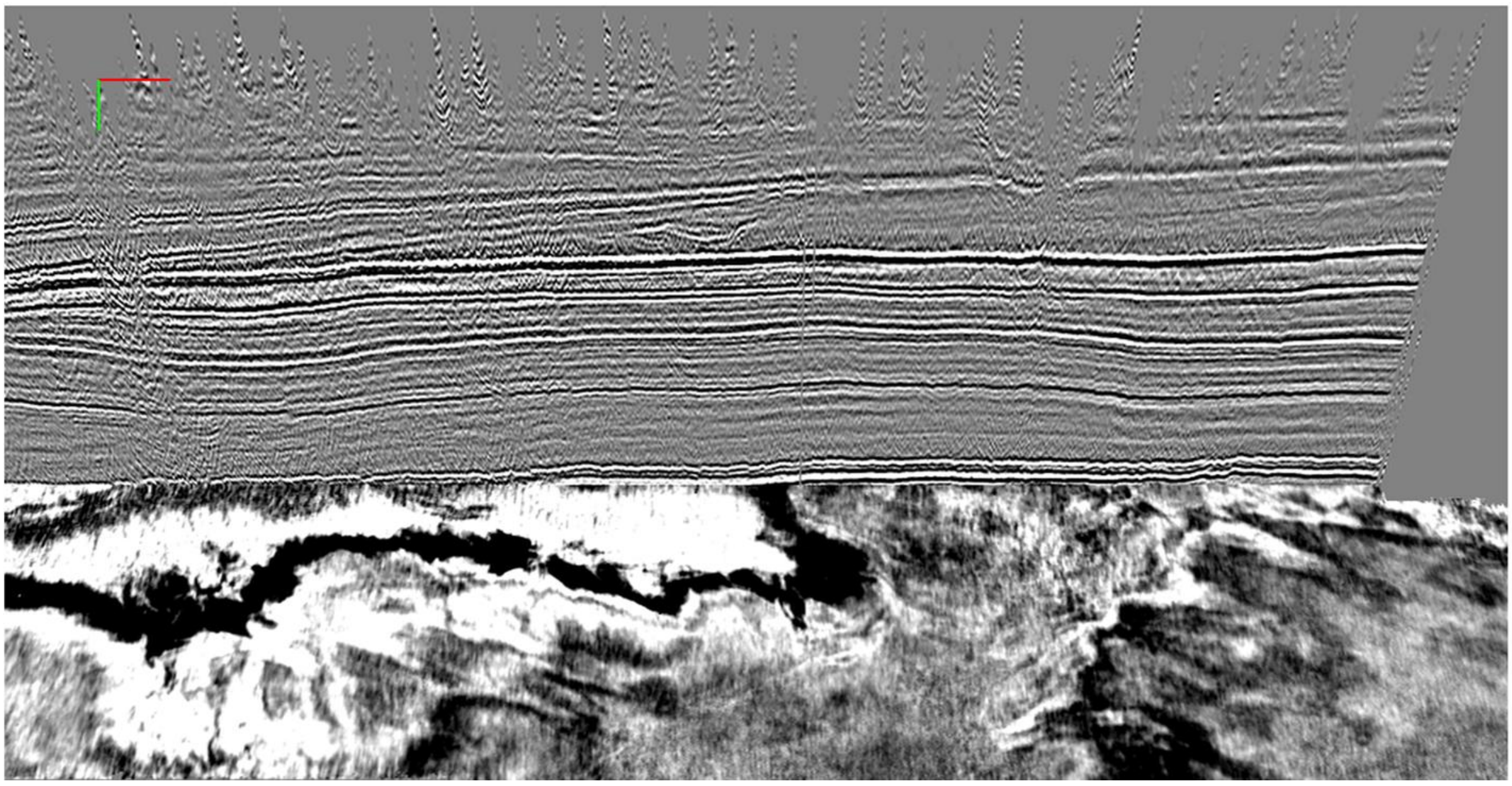

*Fig 9. Dutch onshore 3D seismic cube "Leeuwarden", AI self-supervised denoised northernmost full inline and timeslice 1200 ms. The result is minimal signal leakage and maximum amplitude and phase preservation.*

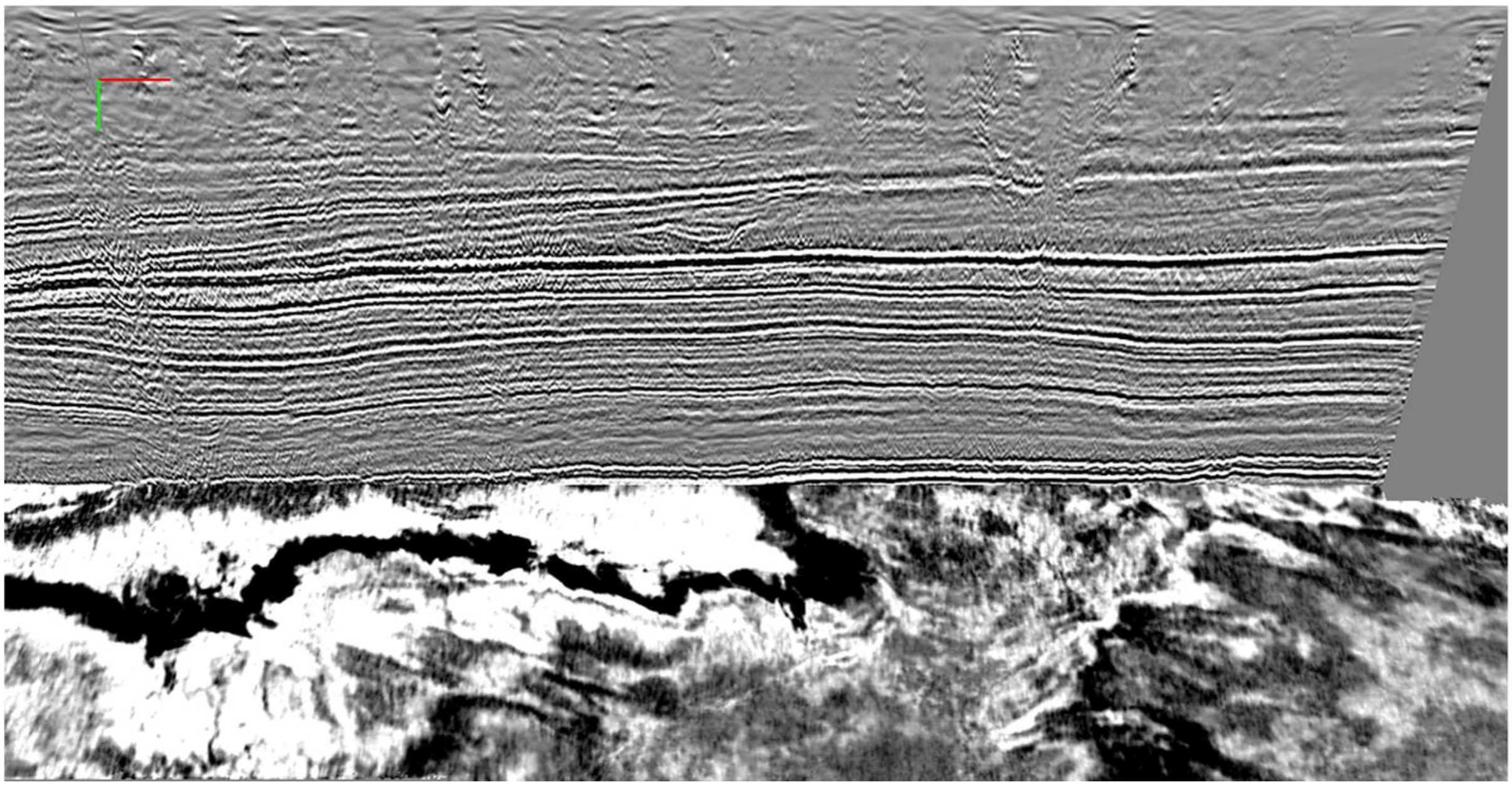

*Fig 10. Dutch onshore 3D seismic cube "Leeuwarden", AI GAN interpolated northernmost full inline and timeslice 1200 ms. The result is maximum gap filling plus amplitude and phase preservation.*

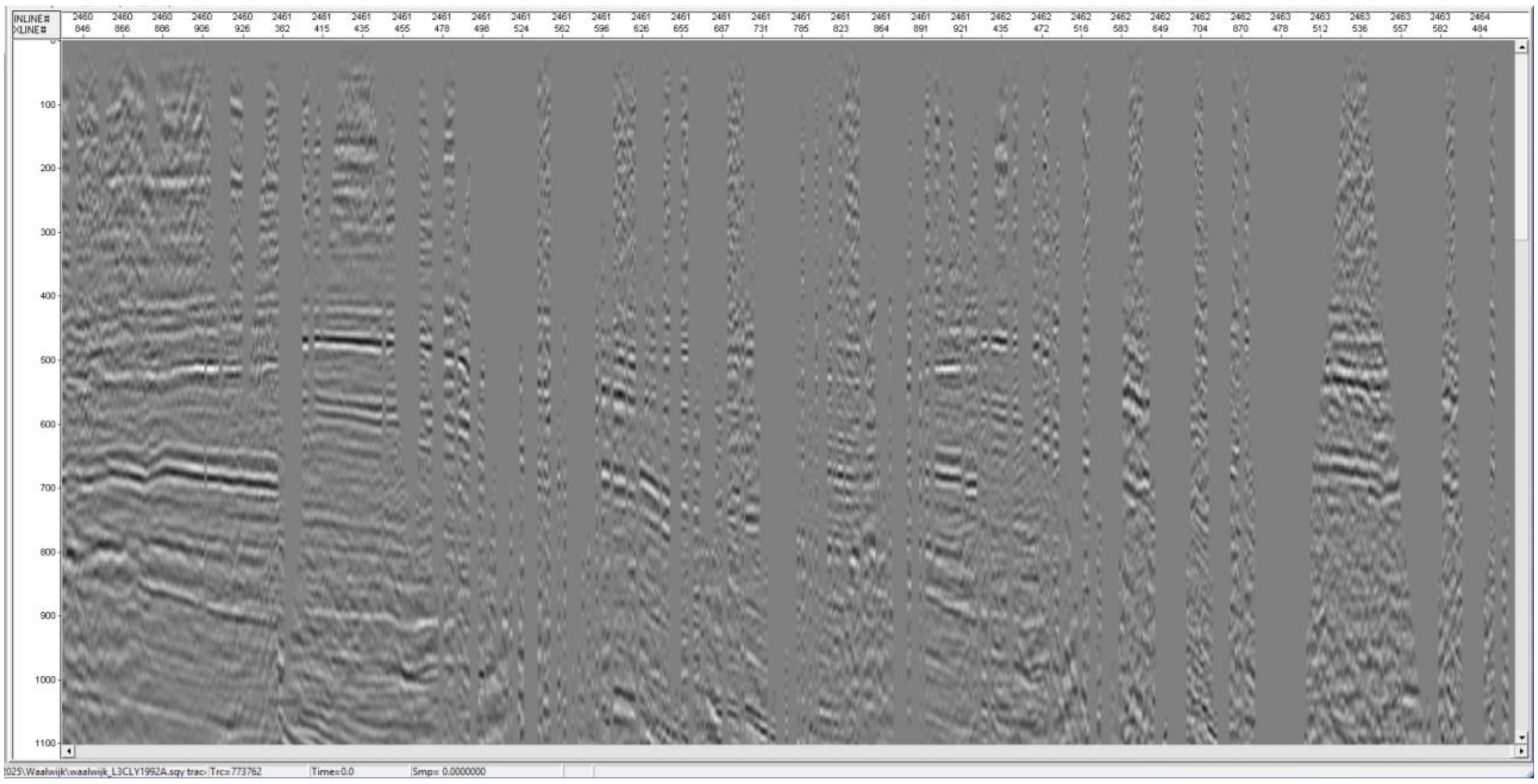

*Fig 11. Dutch onshore 3D seismic cube "Waalwijk", AI GAN interpolated northernmost full inline.*

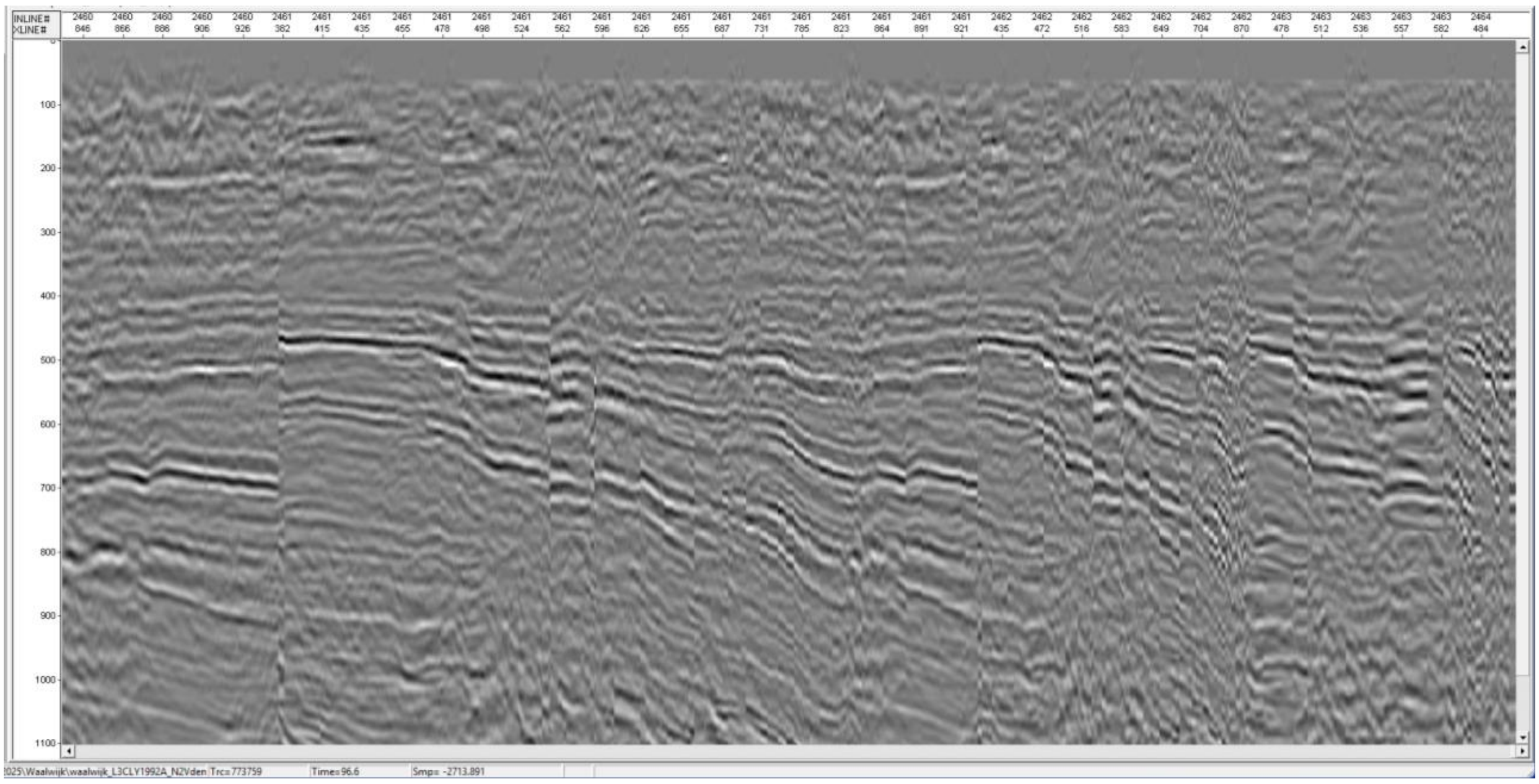

*Fig 12. Dutch onshore 3D seismic cube "Waalwijk", AI GAN interpolated northernmost full inline. The result is maximum gap filling plus amplitude and phase preservation.*

Figures 13, 14, 15, 16, and 17 show the results of the combined workflow: 1) seismic data AI denoising, 2) seismic data AI interpolation, 3) AI interpretation, and 4) human correction of AI interpretation. Figure 13 shows the performance and accuracy metrics of the CONSS self-supervising training phase and Figure 14 shows the training and validation results on a number of inlines in the Leeuwarden 3D seismic cube. A full 3D AI-assited AI interpretation of the "Leeuwarden" conditioned 3D cube (Figure 10) was done with the training data being the DGM Deep and DGM Shallow (NLOG, 2026) interpreted horizons Maassluis, Oosterhout, Breda, and Lower Northsea defined on only 11 inlines. These interpretations in 3D are very consistent and plausible, but the training interpretation of the Maassluis and Oosterhout Formations in particular do not match exactly the local seismic reflections in many places. This is because the input DGM Shallow model (NLOG, 2026) is based solely on borehole data and not on shallow seismic data. Therefore, an interactive "Human-in-the-loop (HITL)" tool has been developed as part of this workflow, which allows human correction of AI interpretations.

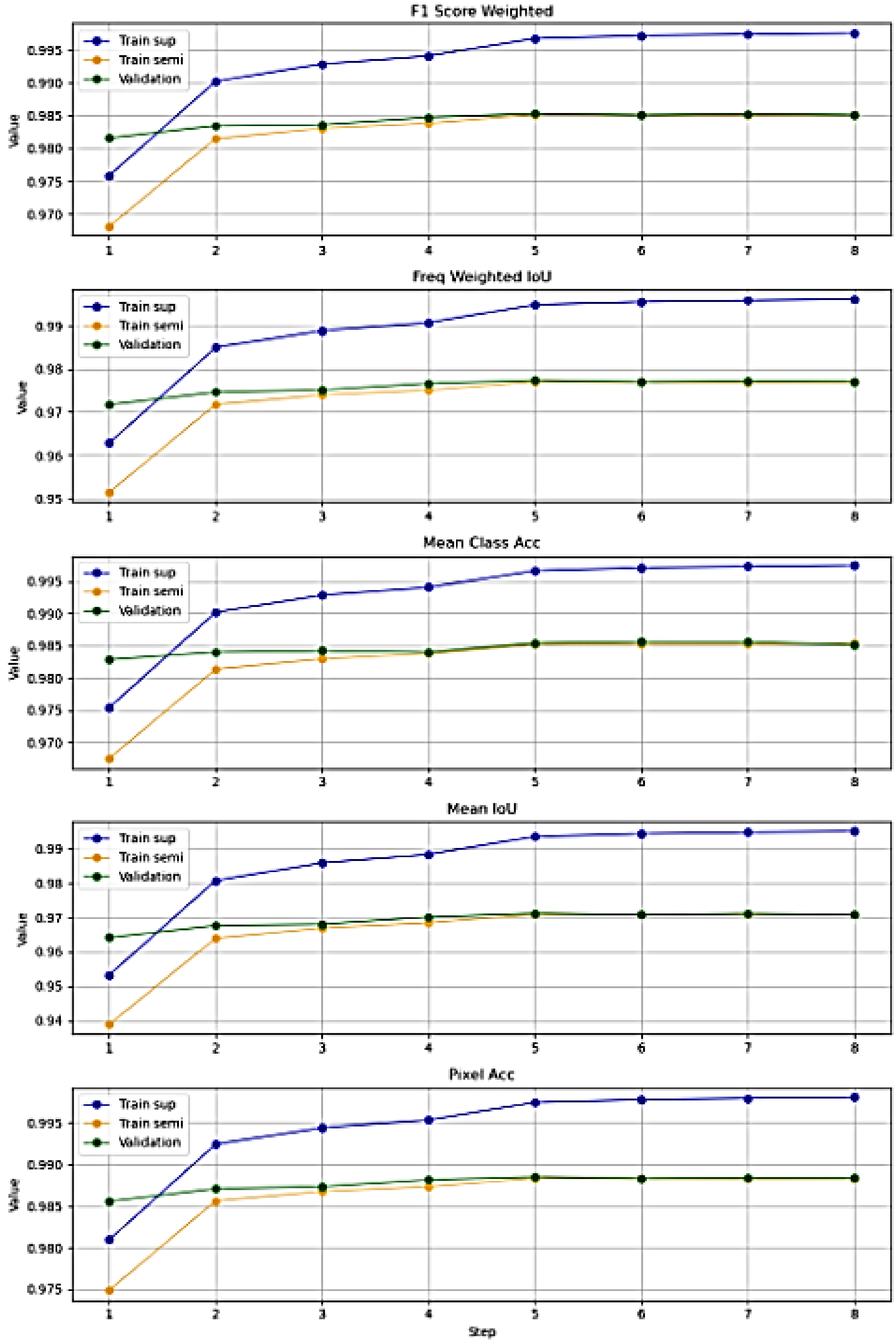

*Fig 13. Metrics of performance of AI tool neural network CONSS on TNO field data. For details on the metrics see Li et al. (2023); higher is better. The blue line represents the metrics for the supervised segmentation head and the orange line represents the metrics for the semi-supervised (contrastive) representation head. The network is first trained to a pretrained model with a lot of general seismic data, to make predictions on new*

*seismic data and retrain the pretrained model to a new model with Contrastive Learning. In this way, a general AI neural network is specifically trained on a new dataset with little required training data.*

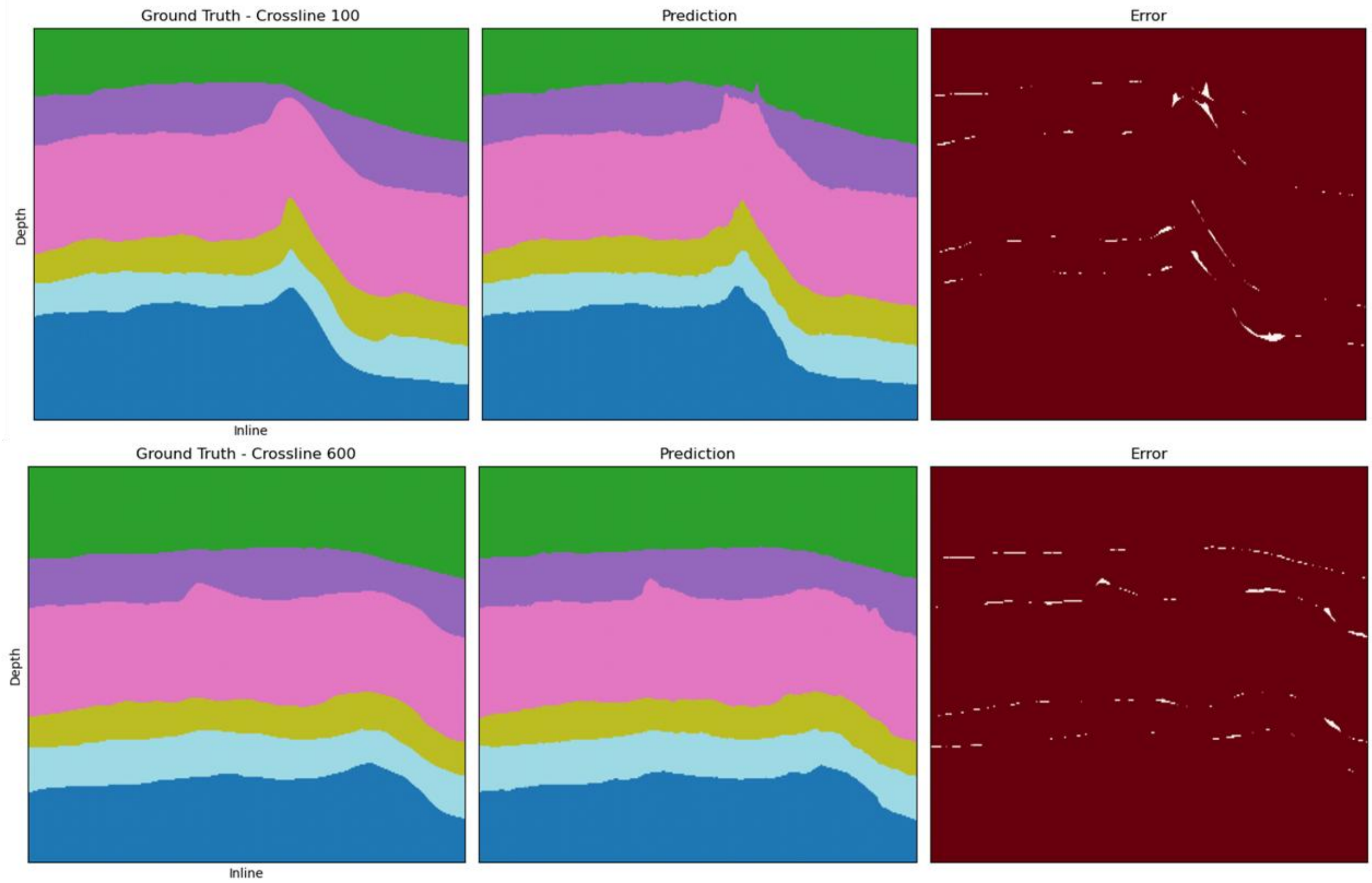

*Fig 14. Comparison of two seismic inline sections ground-truth versus predicted with the AI tool neural network CONSS. For details on the comparison methodology, see Li et al. (2023); lower error (white) is better. The network is first trained with a lot of general seismic data. To make predictions on new seismic data, we retrain the pretrained model with a few representative high-quality lines using Contrastive Learning. This way, a general AI neural network is specifically trained on a new dataset with little required training data and can be used to predict horizons over the entire 3D dataset.*

This line-drawing tool with snap-to-peak can be interactively applied to the old faulty horizon training labels and faulty AI interpretations, and the corrected horizons are retrained with the AI network. The resulting AI-assisted training and prediction on the "Leeuwarden" 3D cube with horizons corrected by this tool can be seen in Figure 15. There, the Top Maassluis (purple) follows the corresponding seismic reflector. In Figure 16, this improved AI interpretation even automatically identified an Elsterian Glacial Valley. Lastly, In Figure 17 a similar training and prediction cycle was performed on the 3D seismic cube "Waalwijk", where highly accurate and plausible clinoforms are predicted in 3D and even an unseen fault is picked up in the middle, cross cutting the shallow Maassluis and Oosterhout strata. This makes the AI-assisted interpretation appear capable of delineating the major stratigraphic units.

Figures 18 and 19 show the results of the workflow: 1) AI interpretation of faults. In these figures, the AI fault detection of the "Waalwijk" 3D cube is performed by a pre-trained model. This pre-trained model was trained by Dou et al. (2023) on synthetic faults and fine-tuned on field seismic data and faults from the F3 cube in the Dutch offshore, defined on only 1 % of the F3 inlines. The fault interpretation in 3D (Figure 18) is consistent and plausible, with faults being detected up to the Maassluis and Oosterhout Formations (Figure 19, shallow up to ± 200 ms), matching the seismic reflections. This makes the AI fault detection workflow a helpful asset for mapping plausible fault networks in both the deep and shallow domain.

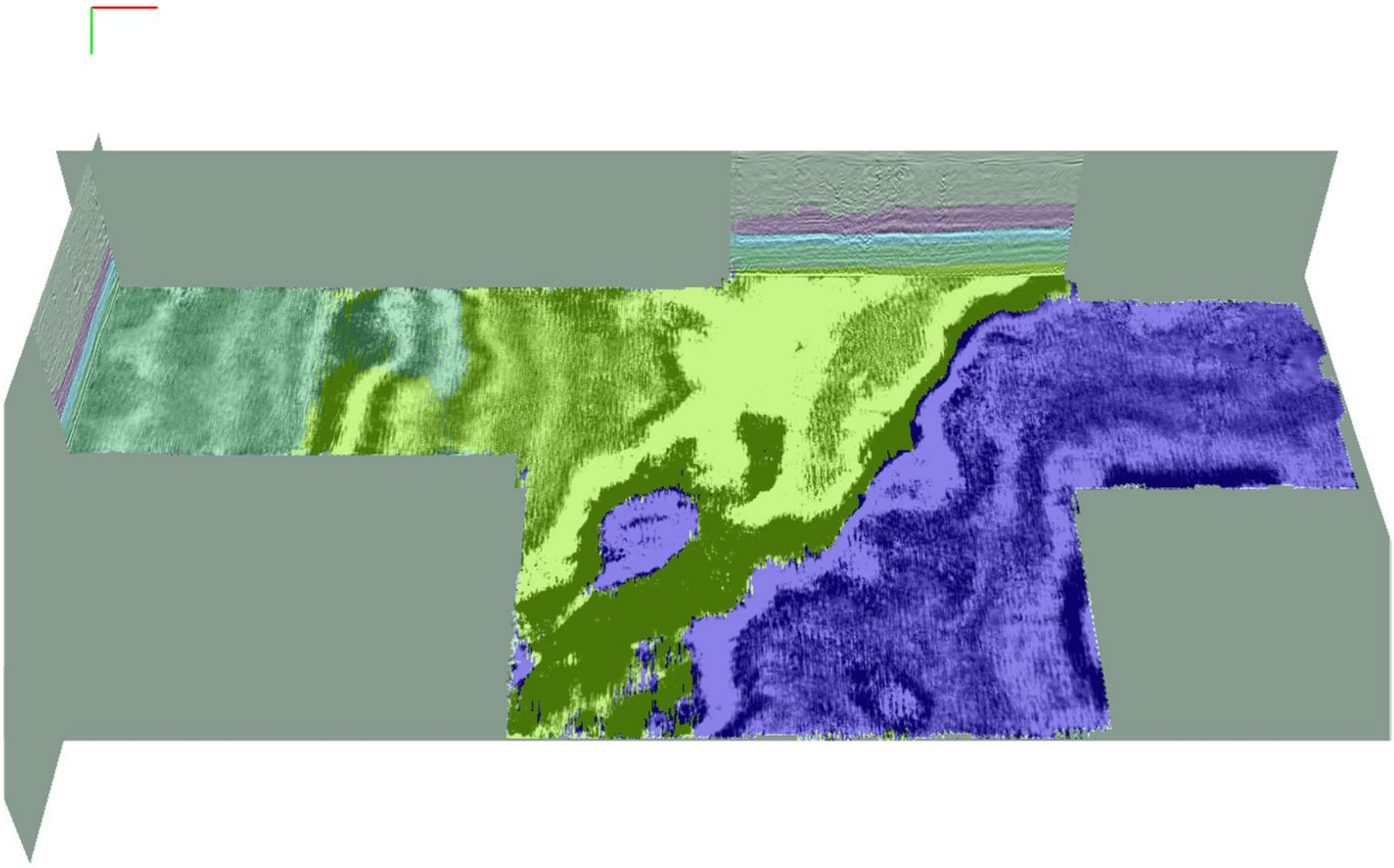

*Fig 15. Dutch onshore 3D seismic cube "Leeuwarden". CNN semi-supervised classified facies on inline and timeslice. Only 1% labeled facies needed for training, 11 inlines in the entire cube. Reinterpreted and re-predicted Top Maassluis. Color codes top-to-bottom: lilac: Maassluis, light blue: Oosterhout, light green: Breda, yellow: Upper North Sea, dark purple: Lower North Sea. Figure orientation and scales: top = north, bottom = south, left = west, right = east; East-west distance = 28 km, north-south distance = 15 km, depth range = 1200 ms (± 1 km).*

This workflow has now been demonstrated on several seismic datasets for the purpose of shallow and deep geothermal exploration. The workflow of AI networks designed appears mature and performs well on synthetic and field seismic data. Training on synthetic data always works well, training on field data is variable because the input labels are not always consistent or correct (sometimes based solely on borehole data) and scale variant. The 3D consistency of the AI predicted geological units, horizons, and faults is plausible, convincing and usable, even in the shallow domain of the Maassluis and Oosterhout Formations.

**Discussion**

The results presented in this study demonstrate that AI-assisted seismic interpretation workflows can substantially improve the efficiency, spatial consistency, and resolution of geological interpretations in shallow to mid-depth onshore settings relevant for geothermal energy and high-temperature aquifer (ATES) applications. By integrating AI-based seismic data conditioning, semi-supervised horizon interpretation, and semi-supervised fault detection into a single workflow, the workflow addresses several long-standing limitations of conventional seismic interpretation in the Dutch subsurface.

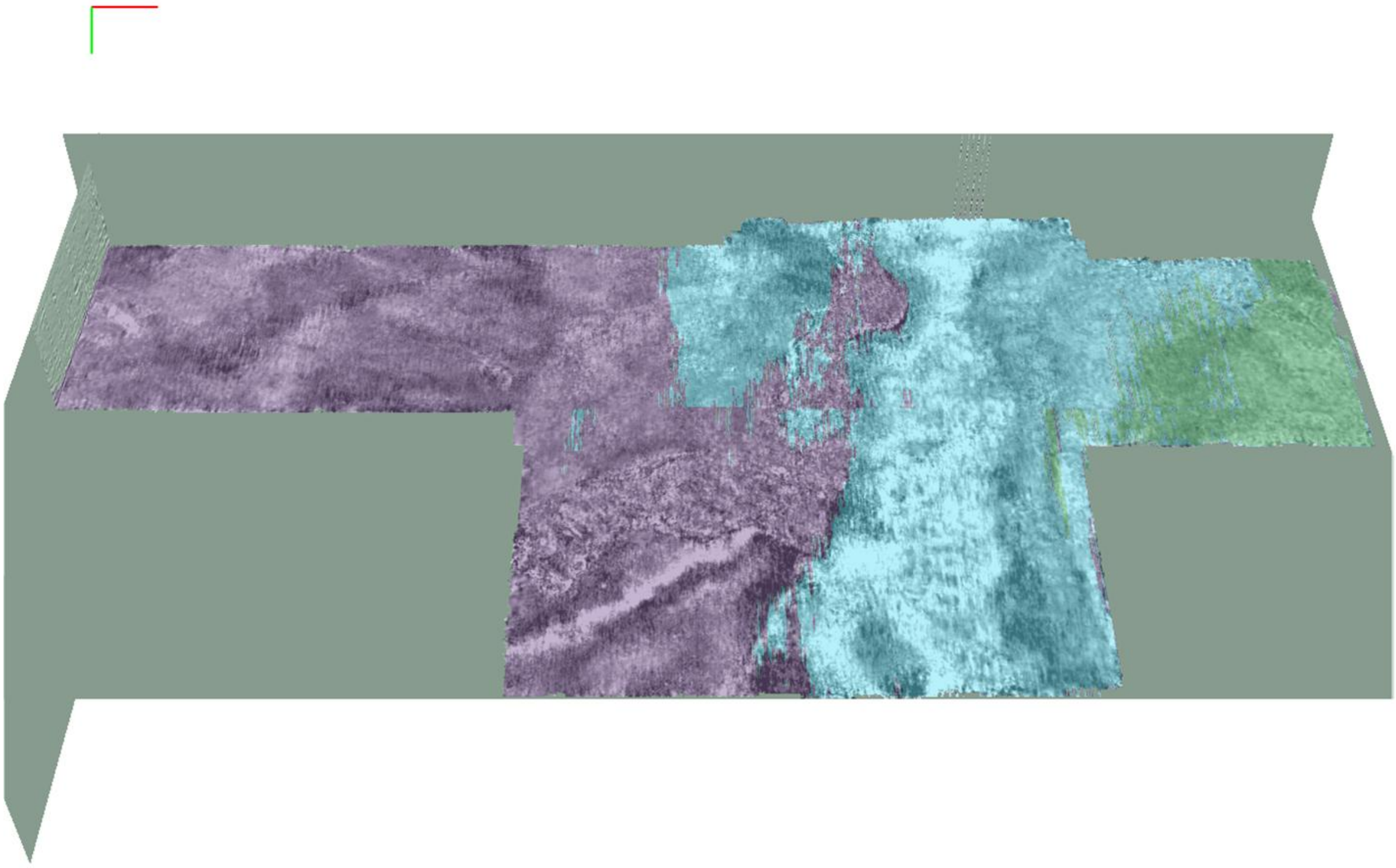

*Fig 16. Dutch onshore 3D seismic cube "Leeuwarden". CNN semi-supervised classified facies on shallower timeslice. Only 1% labeled facies needed for training, 11 inlines in the entire cube. Glacial Elsterian valley recognized, reinterpreted Top Maassluis. Color codes top-to-bottom: lilac: Maassluis, light blue: Oosterhout, light green: Breda. The meandering infill of the (lilac) Maassluis in the light blue Oosterhout represents the Elsterian Valley.*

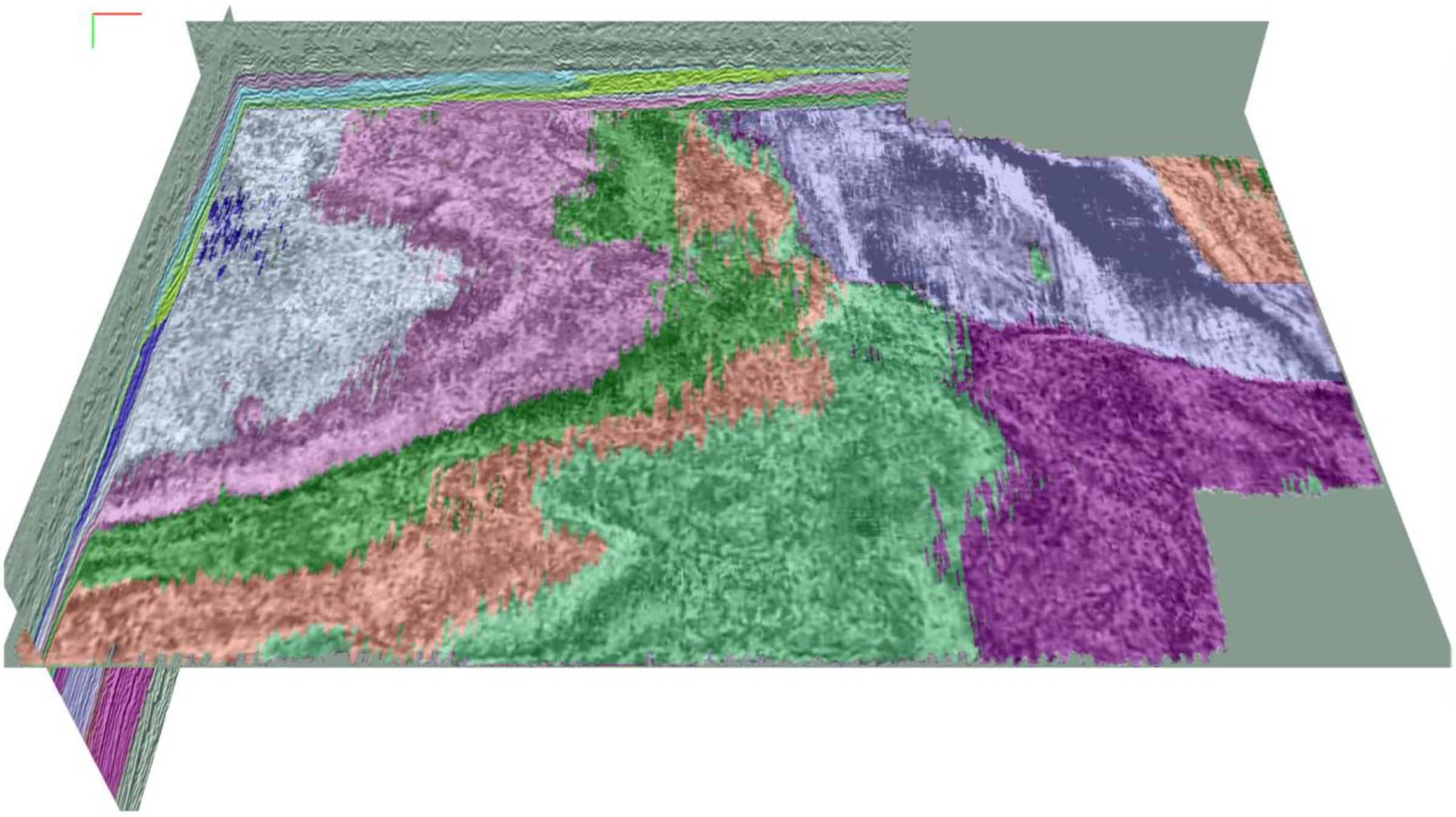

*Fig. 17. Dutch onshore 3D seismic cube "Waalwijk". CNN semi-supervised classified facies on inline. Only 1%*

*labeled facies needed for training, 11 inlines in the entire cube. Prototype interpretation/labeling tool on inline. Color codes from top to bottom: red: Tertiary, light red: Maassluis, light pink: Oosterhout, white: Breda, light blue: Upper North Sea, dark blue: Lower North Sea. Figure orientation and scales: left = west, right = east; East-west distance = 8 km, depth range = 1000 ms (± 0.8 km).*

A key outcome of this work is the demonstrated importance of AI-based seismic data conditioning as a prerequisite for reliable interpretation. Self-supervised denoising and GAN-based interpolation significantly improved reflector continuity and signal coherence in vintage land seismic data, particularly in the shallow domain where noise and acquisition artefacts are most prominent. The improved visibility of the Maassluis and Oosterhout reflections confirms that modern AI conditioning techniques can extract additional geological information from existing datasets that were previously considered of limited value for shallow subsurface characterization. This finding is particularly relevant for regions such as the Netherlands, where new seismic acquisition is constrained and re-use of legacy data is essential. The denoising and interpolation can however not be universally applied, some input-data specific parameters have to be set, like the training data window and post-processing parameters.

The semi-supervised horizon interpretation results highlight the potential of contrastive learning approaches to drastically reduce the amount of required human-labeled training data. Achieving consistent 3D horizon predictions using interpretations from only a small subset of inlines represents a significant advance compared to both manual picking and fully supervised machine-learning methods. However, the results also clearly show that AI predictions remain sensitive to the quality and geological relevance of the training labels. In cases where regional models are based primarily on borehole data and do not fully honor seismic reflectivity, AI predictions may deviate from local seismic signatures. The incorporation of a human-in-the-loop correction step therefore remains essential to ensure geological plausibility and to adapt interpretations to local conditions.

The fault detection results further demonstrate the strength of semi-supervised learning for complex 3D geological features that are difficult to label exhaustively. The FaultSSL approach shows convincing 3D fault continuity and realistic fault geometries, even when trained with limited labeled data. The ability to detect faults consistently up into the shallow Maassluis and Oosterhout formations is particularly important for geothermal and ATES applications, where fault density and connectivity strongly influence reservoir performance and risk assessment. Compared to traditional attribute-based fault detection, the AI-based approach appears less sensitive to noise and processing artefacts, provided that sufficient conditioning and representative training data are available.

Despite these promising results, several limitations and challenges remain. Training on synthetic data consistently yields strong performance, whereas training and fine-tuning on field data show greater variability. This reflects ongoing challenges related to label consistency, scale dependency, and geological ambiguity in real-world datasets. Moreover, while the 3D spatial consistency of AI predictions is generally high, the geological meaning of predicted units and faults must still be validated by experienced interpreters. AI methods therefore should be regarded as powerful accelerators and consistency enhancers rather than replacements for expert geological interpretation.

From an application perspective, the presented workflow provides a scalable and reproducible approach for improving subsurface characterization in data-limited settings. For geothermal energy and ATES projects, where uncertainties in shallow stratigraphy and fault architecture can significantly affect feasibility and public acceptance, the ability to rapidly generate consistent 3D interpretations from existing seismic data is a major advantage. The workflow aligns well with the broader objectives of the WarmingUP GOO project, supporting knowledge development, data re-use, and upscaling of geothermal and subsurface energy applications.

Continued development should focus on improving robustness across varying data qualities, enhancing geological constraints in training datasets, and further integrating AI outputs with implicit geological modeling frameworks. When combined with expert oversight, AI-based interpretation offers a realistic pathway toward faster, more consistent, and more informative subsurface models for the energy transition.

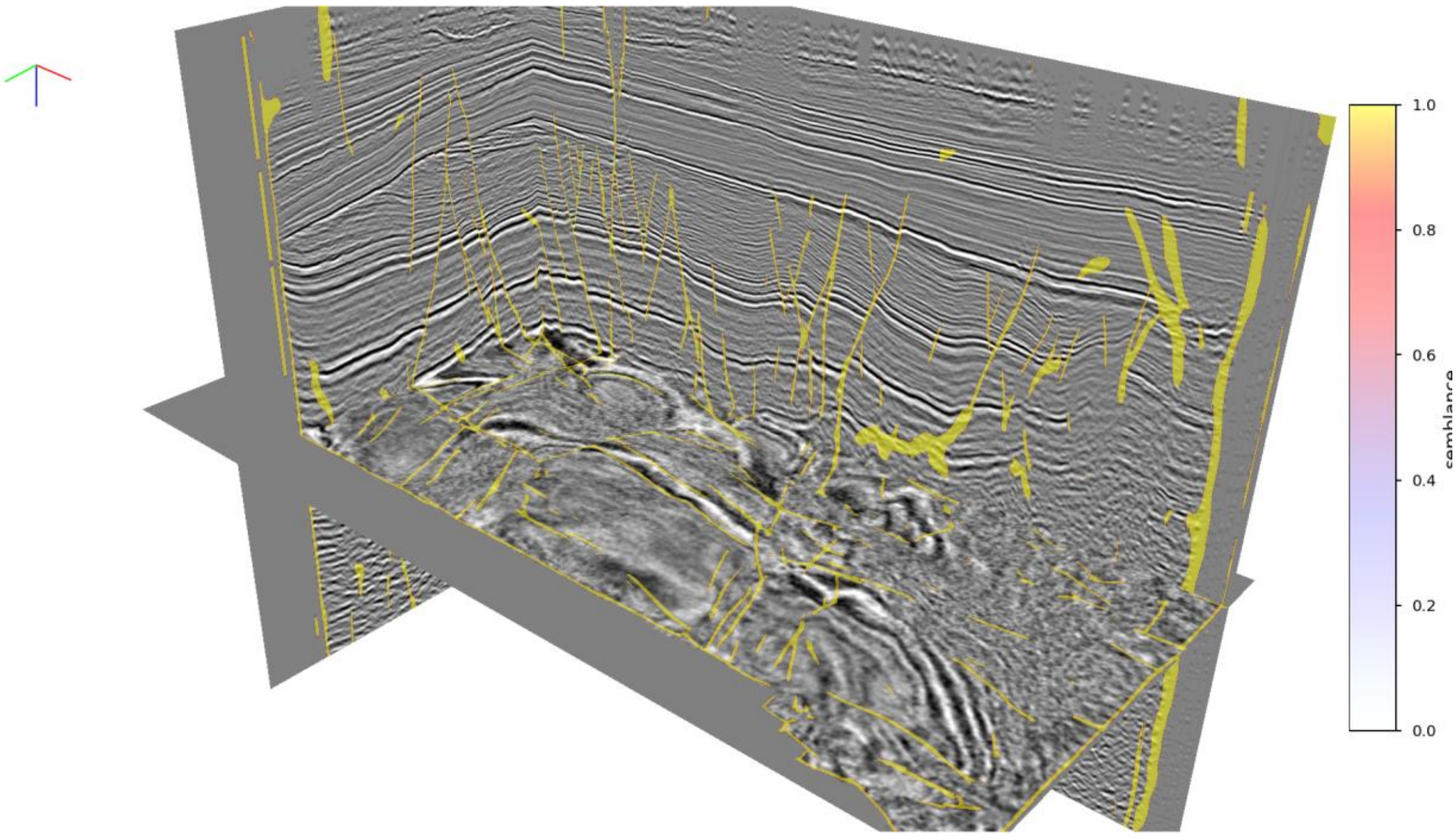

*Fig 18. Dutch onshore 3D seismic cube "Waalwijk", CNN semi-supervised Fault detection, FaultSSL, 3D result, only a small percentage of labeled faults needed, pretrained model, retraining with distilled learning possible. Yellow lines are faults predicted by FaultSSL (probability ± 1) projected on input data Waalwijk.*

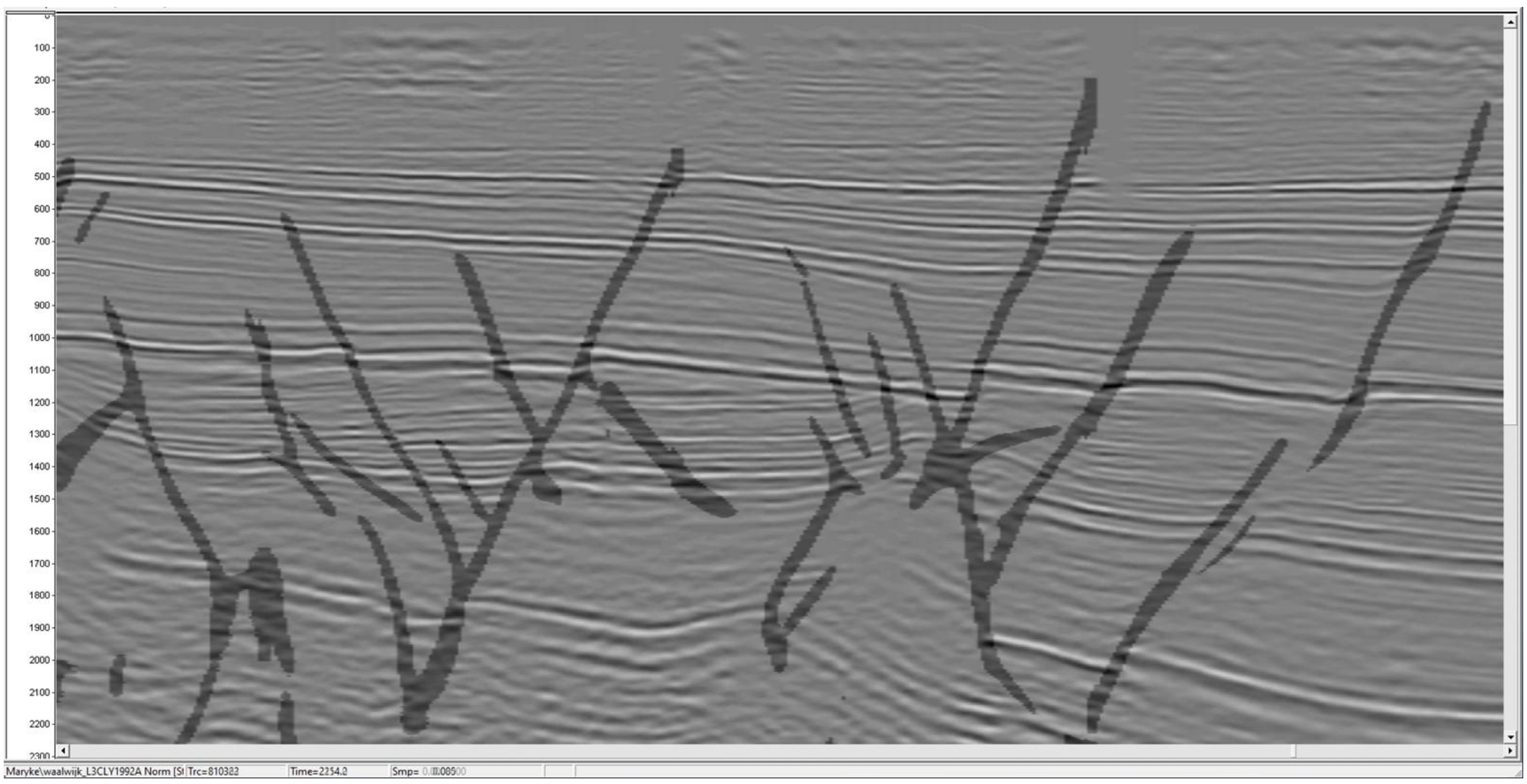

*Fig 19. Dutch onshore 3D seismic cube "Waalwijk", original inline. Superposed with transparent black: FaultSSL predicted faults.*

## Conclusions

This study demonstrates that AI-assisted seismic interpretation workflows can be effectively applied to the characterization of shallow to mid-depth onshore subsurface domains relevant for geothermal energy and high-temperature aquifer (ATES) applications. By combining AI-based seismic data conditioning, semi-supervised horizon interpretation, and semi-supervised fault detection into a single, integrated workflow, the workflow enables rapid, consistent, and scalable interpretation of large volumes of vintage 2D and 3D seismic data.

The results show that self-supervised denoising and GAN-based interpolation are critical enabling steps for improving the quality and interpretability of land seismic data, particularly in the shallow subsurface where noise and acquisition artefacts commonly obscure geological signals. Enhanced reflector continuity and signal coherence directly improve the reliability of subsequent AI-based horizon and fault interpretations, allowing geological features such as the Maassluis and Oosterhout formations to be mapped consistently in three dimensions.

Semi-supervised learning approaches for horizon interpretation and fault detection significantly reduce the dependence on extensive human-labelled training datasets. Consistent 3D predictions were achieved using only a small fraction of manually interpreted inlines, demonstrating a substantial efficiency gain compared to conventional interpretation workflows. At the same time, the study confirms that the geological relevance and consistency of training labels remain a key factor controlling AI performance, particularly when regional models are not fully constrained by seismic reflectivity.

The incorporation of human-in-the-loop correction proves essential for ensuring geological plausibility and for adapting AI predictions to local seismic and stratigraphic conditions. Rather than replacing expert interpreters, the presented workflow positions AI as a powerful tool to accelerate interpretation, enhance spatial consistency, and support more complete use of available subsurface data.

Overall, the workflow has reached a level of maturity that allows its application in applied geothermal and ATES projects, supporting improved subsurface characterization in data-limited settings. By enabling efficient re-use of existing seismic datasets and generating consistent 3D interpretations of horizons and faults, the workflow contributes directly to reducing uncertainty in subsurface models. This provides a strong basis for integration with implicit geological modeling and supports informed decision-making in the context of the energy transition.

**Acknowledgements**

This work was made possible by the Horizon Europe GO-Forward project (grant no. 101147618). Research for this publication has also been made possible with Dutch governmental funding in the MOOI WarmingUP GOO project 60376 and Rijksbijdrage TNO voor Integraal Onderzoeksprogramma, within the demand-driven programs Geo Energy and Geo-Information.